\documentclass[twocolumn,aps,superscriptaddress,nofootinbib,floatfix,linenumbers]{revtex4}
\usepackage{amsfonts}
\usepackage{float}
\usepackage{placeins}
\usepackage{graphicx}
\usepackage[hidelinks]{hyperref}
\usepackage{color,amsmath,amssymb,bm}
\usepackage{tensor}

\usepackage[normalem]{ulem}  

\renewcommand{\sout}{\bgroup \color{red} \ULdepth=-.5ex \ULset}

\def\blfootnote{\xdef\@thefnmark{}\@footnotetext}

\newcommand{\beq}{\begin{equation}}
\newcommand{\eeq}{\end{equation}}
\newcommand{\bea}{\begin{eqnarray}}
\newcommand{\eea}{\end{eqnarray}}

\begin{document}

\title{Probing the electromagnetic fields in ultrarelativistic collisions with leptons from $Z^0$ decay and charmed mesons}

\author{Yifeng Sun}
\email{sunyfphy@lns.infn.it}
\affiliation{Laboratori Nazionali del Sud, INFN-LNS, Via S. Sofia 62, I-95123 Catania, Italy}
\affiliation{Department of Physics and Astronomy, University of Catania, Via S. Sofia 64, 1-95125 Catania, Italy}

\author{Salvatore Plumari}
\email{salvatore.plumari@ct.infn.it}
\affiliation{Department of Physics and Astronomy, University of Catania, Via S. Sofia 64, 1-95125 Catania, Italy}
\affiliation{Laboratori Nazionali del Sud, INFN-LNS, Via S. Sofia 62, I-95123 Catania, Italy}

\author{Vincenzo Greco}
\email{greco@lns.infn.it}
\affiliation{Laboratori Nazionali del Sud, INFN-LNS, Via S. Sofia 62, I-95123 Catania, Italy}
\affiliation{Department of Physics and Astronomy, University of Catania, Via S. Sofia 64, 1-95125 Catania, Italy}

\date{\today}

\begin{abstract}
Ultrarelativistic heavy ion collisions are expected to generate a huge electromagnetic field, $eB \approx 10^{18}\, Gauss$, that induces a splitting of the directed flow, $v_1=\left\langle p_x/p_T \right\rangle $, of charged particles and anti-particles. Such a splitting for charmed meson manifests even for neutral particle/anti-particles pairs ($D^0, \overline{D}^0$), hence being also a unique probe  of the formation of the quark-gluon plasma phase. For the first time here we show that electromagnetic field may  generate a $v_1(D^0) - v_1( \overline{D}^0)$ as huge as the one recently observed at LHC against early expectations.

Within this new research topic we point out a novel measurement: $v_1$ of leptons  from $Z^0$ decay and its correlation to that of $D$ and $B$ mesons. The correlation for $\Delta v_1$ of $l^{+} - l^{-}$, $D^0-\overline{D}^0$ and $B^0-\overline{B}^0$ would provide a strong probe of the electromagnetic origin of the splitting and hence of the formation of a quark-gluon plasma phase with heavy quarks as degrees of freedom. The case of the $v_1({l^{\pm}})$ presents features  due to the peculiar form of the $p_T$ spectrum never appreciated before in the study of heavy ion collisions. We specifically predict a sudden change of the $\Delta v_1(p_T)$ of  leptons at $p_T=45$ GeV$/c$, that can be traced back to a universal relation of  $\Delta v_1$ with the slope of the $p_T$ particle distribution and the integrated effect of the Lorentz force.

\end{abstract}

\maketitle

\section{Introduction} 
The ultrarelativistic heavy ion collisions (uRHICs) at both the Relativistic Heavy Ion Collider (RHIC) and the Large Hadron Collider (LHC) have been a terrestrial tool to access the properties of the Hot QCD matter and
have shown that such a matter expands hydrodynamically with a very small  shear viscosity to entropy density ratio, close to
a conjectured lower limit for strongly interacting matter \cite{Kovtun:2004de}.
At the current stage the description of the expanding dynamics in AA collisions appears to be quite solid which is opening new frontiers
in the study of uRHICs. The main novel aspects that are becoming accessible are those
 related to the impact of the strongest electromagnetic 
field \cite{Das:2016cwd,Chatterjee:2018lsx} and of the largest relativistic vorticity \cite{Becattini:2017gcx,STAR:2017ckg} 
ever created  in a physical system.
This is triggering an intense studies in the noncentral heavy ion collisions, such as the chiral magnetic effect (CME)~\cite{Kharzeev:2007jp,Fukushima:2008xe,Kharzeev:2009fn,Jiang:2016wve,Shi:2017cpu,Sun:2018idn}, chiral magnetic wave (CMW)~\cite{Kharzeev:2010gd,Burnier:2011bf,Yee:2013cya,Sun:2016nig} and the splitting in the spin polarization~\cite{Becattini:2016gvu,Han:2017hdi,Guo:2019joy} and the directed flow ($v_1$) of mesons
~\cite{Gursoy:2014aka,Das:2016cwd,Gursoy:2018yai,Chatterjee:2018lsx}. 

In the last decade it has been possible also to study the dynamics of heavy quarks (HQs), mainly charm quarks, achieving a first phenomenological
determination of their interaction with the quark-gluon plasma (QGP) \cite{Dong:2019unq,Prino:2016cni,Xu:2017obm}. 
This opens the possibility to investigate by mean of HQs new aspects coming from the very early stage dynamics. 
In fact, as pointed out in Ref.~\cite{Das:2016cwd}, charm quark can provide a very suitable tool to explore the presence of the electromagnetic
field thanks to their short formation time given by the $c\overline{c}$ pair production, $\tau_{form} = \frac{\hbar}{2\, m_c} \simeq 0.08 \,\rm fm/c$, that makes them present when the electromagnetic field is also expected to be around its maximum value, while the light quark are likely to appear
later from the decay of the initial glasma phase. Therefore they could be excellent probes of the huge electromagnetic field 
~\cite{Das:2016cwd,Chatterjee:2017ahy,Chatterjee:2018lsx} as recently confirmed by a first measurement
of  $v_1$ of D mesons ~\cite{Adam:2019wnk,Acharya:2019ijj}. 

A breakthrough will be to understand if the splitting very recently observed for $D^0$ ($\rm c\overline{u}$) and $\overline{D}^0$ ($\rm \overline{c}u$) has an electromagnetic origin. 
Here we point out for the first time
that even the measurement by ALICE  \citep{Adam:2019wnk}, much larger and even of opposite sign
than the one predicted in Refs. \cite{Das:2016cwd,Chatterjee:2018lsx}, 
can still be generated by the magnetic field, not with larger maximum strength
but due to a slow damping. We also propose a new measurement in this Letter the $v_1$ as a function of transverse momentum $p_T$ and pseudorapidity $\eta$ of the three generations charged leptons coming from $Z^0$ decay.
Due to the fact that they interact with the electromagnetic field and not with the strong one, they are a more suitable tool
to extract info on the e.m. field, especially because they are generated in the pre-thermal equilibrium stage of QGP
with a decay lifetime very similar to the charm quark formation time which makes them undergo the same electromagnetic field. The absolute measurement of $v_1(p_T,\eta)$  of these  charged leptons $l^{\pm}$ and its correlation to the  $D^0, \overline{D}^0$ will be 
smoking gun of  the electromagnetic origin of the splitting.

\section{Time evolution of electromagnetic fields} 
Though the magnetic field at the moment when colliding nuclei completely overlap $t=0$ can be evaluated with sufficient confidence in relativistic heavy ion collisions~\cite{Skokov:2009qp,Deng:2012pc}, its time evolution is still largely an open question~\cite{McLerran:2013hla,Gursoy:2014aka,Tuchin:2015oka,Inghirami:2016iru}. 
It constitutes a very challenging task the evaluation of the electromagnetic field essentially because it evolves
inside a medium originating under extreme non-equilibrium condition and quickly evolving from glasma matter
to quark-gluon matter. Furthermore the value of conductivity for Hot QCD matter in lattice QCD has still large 
uncertainties  \cite{Ding:2010ga,Amato:2013naa}. In addition, an anomalous conductivity can be generated by the formation of chiral topological charges~\cite{Tuchin:2019gkg}.
To have a general study of the effects of electromagnetic fields spanning over different profiles currently
employed, we include three typical configurations, 
where only $E_x$ and $B_y$ are included due to their dominance over other contributions. 

\textit{Case A} - A standard approach, developed in several papers ~\cite{Tuchin:2013apa,Gursoy:2014aka,Das:2016cwd}, has been based on the solution
of the Maxwell equations for the field generated by the spectators in the overlapping region assuming this is made by
matter in equilibrium with a constant conductivity $\sigma_{el}=0.023$ fm$^{-1}$ within lQCD 
calculations~\cite{Ding:2010ga,Amato:2013naa,Brandt:2012jc}. The drawback
is to assume a finite conductivity associated to the QGP even before the collisions, hence strongly damping 
the maximum value of $B_y$ ($\sim 50$ times)  \cite{Deng:2012pc}
w.r.t. the vacuum estimate.

\textit{Case B} - The large gap between the maximum initial $B_y(t=0)=B_0$ field with and without assuming  a conducting medium
has led other authors to consider a parametrization that at least at initial time agrees with the large value expected
in the vacuum. 
We label this case as B:
$eB_y(x,y,\tau)=-B(\tau)\rho_B(x,y)$ with $\rho_B(x,y)=\rm{exp}[-\frac{x^2}{2\sigma_x^2}-\frac{y^2}{2\sigma_y^2}]$~\cite{Roy:2017yvg}
and 
$B(\tau)=eB_0/(1+\tau^2/\tau_B^2)$, which represent the transverse coordinate dependence and time evolution of $B_y$, as developed in Refs~\cite{Jiang:2016wve,Shi:2017cpu} with $B_0$ given by
 in vacuum estimate.
 $eE_x$ is then determined by solving the Faraday's Law $\boldsymbol{\nabla}\times\mathbf{E}=-\partial \mathbf{B}/\partial t$:
\begin{eqnarray}
eE_x(t,x,y,\eta_S)=\rho_B(x,y)\int_0^{\eta_S}d\chi B^{'}(\frac{t}{\rm{cosh}\chi}) \frac{t}{\rm{cosh}\chi}
\end{eqnarray}
where $\eta_S$ is space-time rapidity. In the above, $\tau$ and $\eta_S$ are related to $t$ and $z$ by $\tau=\sqrt{t^2-z^2}$ and $\eta_S=\frac{1}{2}\rm ln(\frac{t+z}{t-z})$.
We will see that such a field even if much larger than case A still does not lead to a prediction in agreement with the
large splitting observed by ALICE at LHC \cite{Adam:2019wnk}, indeed it even leads to a 
$\Delta v_1=v_1(D^0)- v_1(\overline{D}^0)$  of opposite sign.

\textit{Case C} - A third case for the magnetic field $B_y$ starts from the same initial value
and the space distribution of case B, but has a slower time evolution of the field: $B(\tau)=eB_0/(1+\tau/\tau_B)$.

\begin{figure}[h]
\centering
\includegraphics[width=1\linewidth]{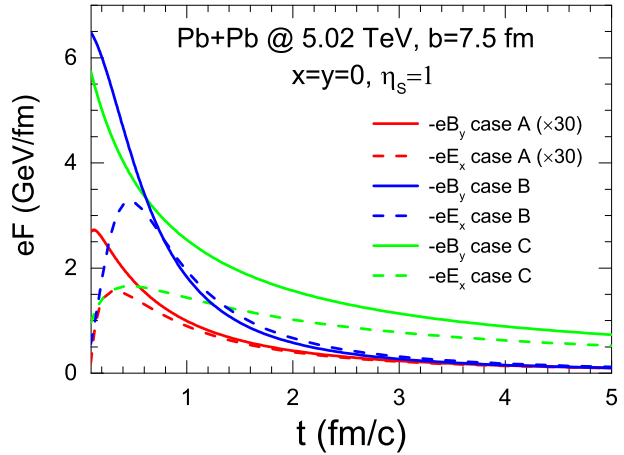}
\caption{(Color online) 
Time evolution of $-eB_y$ and $-eE_x$ at $x=y=0$ and space-time rapidity $\eta_{S}=1$ in 5.02 TeV Pb+Pb collision at impact parameter $b=7.5$ fm for three cases. See text for details.}
\label{fig:eF}
\end{figure}

In this study we focus on 5.02 TeV Pb+Pb collisions at 20-30\% centrality, that corresponds to impact parameter $b=7.5$ fm, due to the recent measurements by ALICE of non-zero directed flow splitting between $D^0$ and $\overline{D}^0$~\cite{Acharya:2019ijj}. 
Though we choose this system to illustrate the effect, our conclusions are pretty general. In Fig.~\ref{fig:eF} we show the time evolution of $eB_y(t)$ and $eE_x(t)$ at $x=y=0$ and $\eta_S=1$ for the three cases discussed above. $eB_y$ is chosen to be negative 
as in the experimental convention.

For cases B and C, the parameters are found to be $eB_0=73\, m_{\pi}^2$, $\sigma_x=3$ fm and  $\sigma_y=4$ fm. 
The $\tau_B$ has a large uncertainty but we fix it to be the same for case B and C with
$\tau_B$=0.4 fm$/c$. 
It is seen that in case B the time derivative of the $B_y(t)$ is such to generate a $E_x \simeq B_y$ at $t \simeq \, 0.6 \, \rm fm/c$, while for case C the slower time evolution leads to a quite smaller $E_x$ that essentially remains always smaller than $B_y$, even if the initial values of case B and C are equal and  remain so up to time $t \approx 1\, \rm fm/c$.

\section{Spectra of leptons and charm quarks} 
\begin{figure}[h]
\centering
\includegraphics[width=1\linewidth]{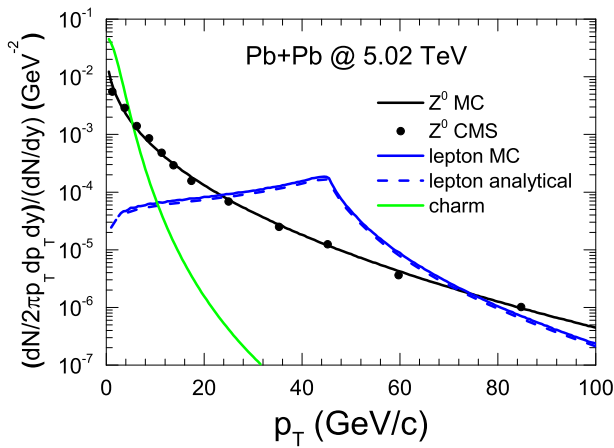}
\caption{(Color online) 
Normalized spectra of $Z^0$ boson and lepton as a function of transverse momentum $p_T$ from Monte Carlo simulation in 5.02 TeV Pb+Pb collisions, compared to the CMS measurements~\cite{Khachatryan:2015pzs} of $Z^0$ and an analytical calculation of lepton from $Z^0$ decay. The green line corresponds to the spectrum of charm quarks when they are initially formed by pair production.}
\label{fig:spectra}
\end{figure}

We use the experimental measurements~\cite{Chatrchyan:2014csa,Khachatryan:2015pzs} to construct a parametrization of the transverse momentum and rapidity dependence of $Z^0$.
 It is seen in Fig.~\ref{fig:spectra}  that $Z^0$ generated by a Monte Carlo simulation with such parametrization does agree quite well with the CMS measurements shown by black dots~\cite{Chatrchyan:2014csa}. The distribution in the transverse plane of $Z^0$ is given 
 by the binary collisions of colliding nuclei, while in the longitudinal axis $z=\tau_{Z^0}\rm{sinh}y_z$ and $t=\tau_{Z^0}\rm{cosh}y_z$ with $\tau_{Z^0}=\hbar/m_{Z^0}=0.0022$ fm$/c$, where $y_z$ is the rapidity of particles.

In Fig.~\ref{fig:spectra} for the leptons decayed from $Z^0$ with mean decay proper time $1/2.495$ GeV$^{-1}$, we use both analytical and Monte Carlo methods to generate their spectra. It is seen that the spectra of leptons peaks at about 45 GeV$/c$, which is due to the kinematic effect.

We also show the transverse momentum distribution of charm quarks initially produced at 5.02 TeV Pb+Pb collisions by the green line in Fig.~\ref{fig:spectra}, which is obtained by the Fixed Order+Next-to-Leading Log (FONLL) QCD~\cite{Cacciari:2005rk,Cacciari:2012ny}. It is seen that the distributions of charm quarks and leptons from the  decay of $Z^0$ take quite different $p_T$ dependences, which we show later that this can lead to different patterns on the $p_T$ dependence between the $v_1$ of leptons and charm quarks.

\section{Numerical results} 
We study the evolution of both lepton and $D^0$ meson by standard Langevin equations~\cite{Gossiaux:2008jv,Gossiaux:2009mk,Cao:2015hia,Xu:2017obm,vanHees:2005wb,vanHees:2007me,He:2011qa,Alberico:2011zy,Alberico:2013bza} 
but including Lorentz force~\cite{Das:2016cwd,Chatterjee:2018lsx}:
\begin{eqnarray}
&&d x_i = \frac{p_i}{E}dt,
\label{velocity}\\
&&{d p_i} = -\Gamma p_idt+\xi_{i}\sqrt{2D_pdt}+q(E_i+\epsilon_{ijk}v_jB_k)dt,
\label{force}
\end{eqnarray}
where the momentum diffusion coefficient $D_p$ is related to the drag coefficient $\Gamma$, energy of charm quarks $E$, and the local temperature $T$ by $D_p=\Gamma ET$, and $\xi_i$ is a real number randomly sampled from a normal distribution with $\langle\xi_i\rangle=0$ and $\langle \xi_i\xi_j\rangle=\delta_{ij}$. In addition, $i$ represents the index for the 3D coordinate and momentum, and $v$ is the velocity of the charged particles in the above. For charm quarks, the drag coefficient is determined by 
the study of the D meson observables ~\cite{Scardina:2017ipo,Greco:2017rro,Plumari:2015cfa,Plumari:2019gwq,Sun:2019gxg}. 
We instead include only Lorentz force in Eq.~(\ref{force}) for leptons from $Z^0$ decay due to their negligible interaction with QGP.

\begin{figure}[h]
\centering
\includegraphics[width=1\linewidth]{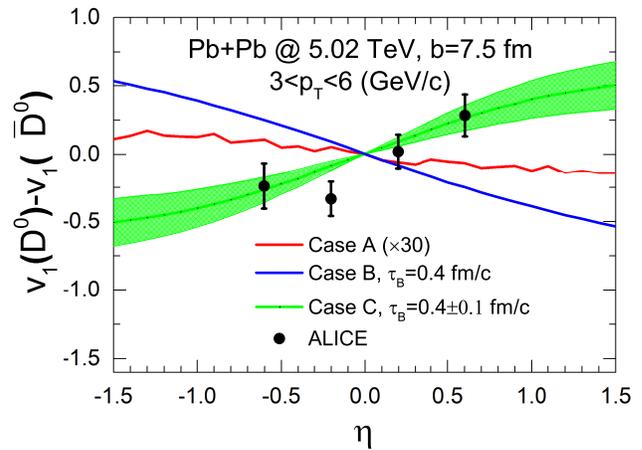}
\caption{(Color online) 
$v_1(D^0)-v_1(\overline{D}^0)$ as a function of pseudorapidity for three cases and compared to ALICE measurements~\cite{Acharya:2019ijj} in 20-30\% centrality Pb+Pb collisions.}
\label{fig:splitting}
\end{figure}

In Fig.~\ref{fig:splitting}, we show the directed flow splitting between $D^0$ and $\overline{D}^0$ as a function of pseudorapidity  within range of $3<p_T<6$ GeV$/c$. It is found that $d\Delta v_1/d\eta$ is -0.004 for case A, -0.42 for case B with $\tau_B=0.4$ fm$/c$, and $0.44 \pm 0.15$ for case C with the interval
of values corresponding to $\tau_B=0.4 \pm 0.1 \rm \, fm/c$. Compared to ALICE measurements~\cite{Acharya:2019ijj} of $d\Delta v_1/d\eta=0.49\pm0.18$, we find that only case C with $\tau_B$ in the range of 0.3-0.5 fm$/c$ can lead to a correct prediction.
In particular, it is surprising that for case B and C even starting with the same initial values of $B_y$ 
and similar magnitude up to $t \approx 1 \, \rm fm/c$ one can have completely opposite
predictions for $d\Delta v_1/d\eta$.

Such a feature is however not accidental, and 
even if this is the result of a quite involved dynamics we can catch the key features determining the above result
by mean of suitable approximations. To the end of deriving analytical formulas
that catch the essential features, we assume that charm-bulk interaction does not affect the splitting
and assume also that the electromagnetic term $qB_0\tau_B$  is a small perturbation
of the charm quark (or lepton) energy. We look for the shift in $p_x$ of a charm quark (lepton) 
of momentum $p_T$, $y_z$ at the initial transverse position $\vec r_0$.
The average value of it  comes from the integral over the participant space region of the $\Delta p_x(\vec r_0)$: 
$\overline{\Delta p}_x(p_T,y) = \int d^2 r_0 \rho(\vec r_0) \Delta p_x(\vec r _0, p_T,y)$,
where $\rho(r_0)$ is the initial distribution of the particles in transverse plane at the formation time $t_0$.
Assuming that $y_z=\eta_s$ and evaluating the shift of each particle as the propagation along straight lines from Eq. (\ref{velocity}), we can write:
$\Delta  p_x(\vec {r}_0,p_T,y_z)=  \int dt q[E_x(t, \vec r(t),y_z)- v_z B_y(t,\vec r(t),y_z)]\nonumber$,
where  $\vec r (t)= \vec r_0 + \Delta \vec r (t-t_0)$ gives the position of the particle at time $t$.
One can write the average $p_x$ shift of the particle more explicitly as:
\begin{eqnarray}
&&\overline{\Delta  p}_x(p_T,y_z)=\int_{t_0}^{\infty}dt \int dx_0 dy_0 \, \rho(x_0,y_0) \int \frac{d\phi}{2\pi}\nonumber
\\&&q \left[ \rm{tanh} y_z \,B \left(t, y_z \right) +\int_0^{y_z} d\chi \,B^{'}\left(t, \chi \right) \frac{t}{\rm{cosh}\chi}\right]\nonumber
\\&&\rho_B \left[  x_0+\frac{p_T\rm{cos}\phi }{m_T\rm{cosh}y_z} (t-t_0), y_0+\frac{p_T\rm{sin}\phi }{m_T\rm{cosh}y_z}(t-t_0)\right] 
\label{change2}
\end{eqnarray}

We can find some general features of the integration over $x_0, y_0$ and $\phi$ without knowing the forms of $\rho(x_0,y_0)$. The integration depends only on the factor $\gamma=\frac{p_T}{m_T}(\frac{t} {\rm{cosh}y_z}-\tau_0)$ with $\tau_0=t_0/\rm cosh y_z$,  and it should be quite uniform in the region of $\gamma<R\sim\sqrt{\sigma_x\sigma_y}$ and decreases fast outside. Because of these features, we can replace the integration 
by a step function $K \Theta(1-\gamma/R)$, where $K$ is some constant depending on the specific forms of $\rho$ and $\rho_B$. Hence integrated by parts, Eq.~(\ref{change2}) can be further simplified to:
\begin{eqnarray}
\overline{\Delta  p}_x(p_T,y_z) \propto q\int_0^{y_z}\frac{d\chi}{\rm{cosh}\chi}\left[ \tau_2B(\tau_2)-\tau_1B(\tau_1)\right] 
\label{final}
\end{eqnarray}
with $\tau_{1}=\frac{\tau_0\rm{cosh}y_z}{\rm{cosh}\chi}$ and $\tau_{2}=\frac{(\tau_0+Rm_T/p_T)\rm{cosh}y_z}{\rm{cosh}\chi}$, 
and $\tau_{1,2}$ can be treated as the formation time and the escape time out of the electromagnetic field of the particle.

Using the form of $B(\tau)$ from case B and C and $\tau_0\sim$0.1 fm$/c$, we can find that $\overline{\Delta p}_x (y_z>0)$ of charm quarks is positive for case C and negative for case B and so will be $\Delta v_1$. One can also find $\overline{\Delta p}_x (y_z>0)$ is negative for case A as well, though with a much smaller magnitude.
In the full calculation one should take into account the interaction with the medium (for charm quarks)
and the not small $qB_0\tau_B$, that induce some further modulation w.r.t. Eq.~(\ref{final}) for low $p_T$ particles. Moreover, we find that with a reasonable interaction strength~\cite{Scardina:2017ipo,Greco:2017rro,Plumari:2015cfa,Plumari:2019gwq,Sun:2019gxg}, the suppression of $\Delta v_1$ of charm quark by medium interaction is negligible at $p_T>$ 3 or 4 GeV$/c$. Eq. (\ref{final}) can thus catch the main feature and predict correctly the sign and even the magnitude with reasonable accuracy, 
as the results of the simulations in Fig.~\ref{fig:splitting} show. 
Eq. (\ref{final}) enlightens our understanding showing that the sign of $\Delta v_1$ is not determined by the initial 
maximum value of $eB_0$, but by its time evolution. It includes implicitly the effect of $E_x$ 
directly related to the time derivative of $B_y$: a slower time evolution leads to a smaller electric field $E_x$, 
inducing a positive sign of $\Delta v_1$. Moreover,  $\Delta v_1$ is actually determined by the variation of $\tau B(\tau)$ reading from Eq. (\ref{final}). This is very relevant and general to orientate the investigations that will grow in this new subject.
In the following, we will focus our attention on the proposal of a new measurement: the $v_1$ of the leptons from
$Z^0$ decay.  We will consider case C that is the only one able to account for the observed positive splitting
of $\Delta v_1^{D} = v_1(D^0) - v_1( \overline{D}^0)$. 

\begin{figure}[h]
\centering
\includegraphics[width=1\linewidth]{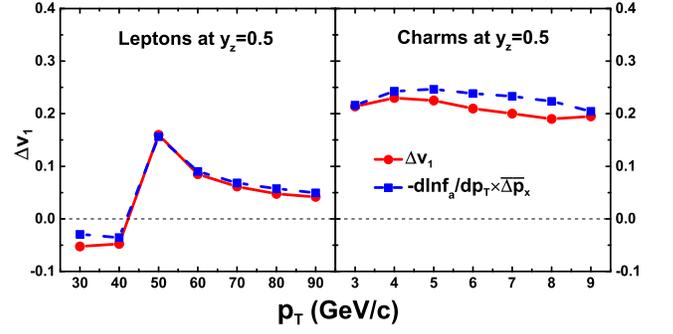}
\caption{(Color online) $v_1$ splitting of leptons from $Z^0$ decay (left panel) and charm quarks (right panel)
as a function of  $p_T$ at 5.02 TeV 20-30\% centrality Pb+Pb collisions of case C; by blue solid squares the predictions
according to Eq.(\ref{v1}).
}
\label{fig:v1pt}
\end{figure}

In the left of Fig. \ref{fig:v1pt} we show the splitting $\Delta v_1^{l}= v_1(l^+) - v_1(l^-)$ of leptons  for case C. It is seen that $d\Delta v_1^l/d\eta$ is negative at $p_T \sim 40 \rm\, GeV/c$, hence
opposite to the one of the charm quarks which is already not trivially expected. Even more relevant is the observation
that  $d\Delta v_1^l/d\eta$  changes abruptly sign around $p_T\sim 45 \,\rm GeV/c$ reaching a peak value comparable to
the one observed and here predicted for  $d\Delta v_1^D/d\eta$ in the region $p_T \sim 3-10 \,\rm GeV/c$.
This sign change for $l^{\pm}$ is quite surprising and not expected if one
relates the flow to the direct effect of the Hall and Faraday drift on one single particle.
In this Letter we indeed clarify that  differential $\Delta v_1$ is not a direct measure of $\overline{\Delta  p}_x$, 
but relates also to the spectra of  particles. 
In the specific case of  the lepton from $Z^0$ decay one has unique shape of the spectrum,
very different from the particle spectra of hadrons in uRHICs, as shown in Fig.\ref{fig:spectra}.
This induces a new feature in the directed flow never anticipated before.


Electromagnetic fields move charged particles from initial $p_{xi}$ to final 
$p_{xf}=p_{xi}+\Delta_x$, where $\Delta_x$ takes a distribution $\rho(\Delta_x)$ around the average $\overline{\Delta p}_x(p_{Ti},y_z)$. For $p_T$ much larger than
the width and average of $\rho(\Delta_x)$ the final momentum distribution can be expanded as:
\begin{eqnarray}
\frac{dN}{dp_xdp_y}
\approx
f_a(p_T)-\overline{\Delta p}_x(p_T,y_z)\frac{\partial f_a}{\partial p_T}\frac{p_x}{p_T},
\label{spectra}
\end{eqnarray}
where $f_a$ is the initial distribution of charged particles, $a=c, \bar c, l^+,l^-$ and we know from the simulation that 
the average value of $\overline{\Delta p}_x$ is about 0.36 GeV/c for charm quarks and 0.7 GeV/c for leptons at $y_z=0.5$
with a weak dependence on $p_T$. 
We thus further find that the differential $v_1(p_T,y_z)$ will be:
\begin{eqnarray}
&&v_1(p_T,y_z) \approx 
\frac{\overline{\Delta p}_x(p_T,y_z)}{2}\frac{-\partial {\rm{ln}} f_a}{\partial p_T}.
\label{v1}
\end{eqnarray}

The goodness of the approximation in Eq (\ref{v1}) is shown 
in Fig. \ref{fig:v1pt} where the splitting $\Delta v_1^a(p_T,y_z)$ is shown for  $a=l$ on the left panel and $a=c$
on the right panel. 
The comparison between red solid circles and blue squares shows how Eq. (\ref{v1})
is able to get the main behavior of the splitting for both charm quarks and leptons from $Z^0$ decay.
Notice therefore the $\Delta v_1^l$ is not trivially related to the D meson one, in particular a small
negative $\Delta v_1^l$ up to $p_T \sim 40$ GeV will not be in contradiction with a large positive $\Delta v_1^D$,
a feature not expected but that is enlightened by Eq. (\ref{v1}).

Due to the similar formation time of leptons from $Z^0$ decay and charm quarks, $\overline{\Delta p}_x$ for these two kinds of charged particles should differ mainly by their different charge. Moreover, because the Lorentz force is not sensitive to $p_T$ for $p_T\gg m$, $\overline{\Delta p}_x$ will be independent of $p_T$ at high $p_T$. These features thus provide a strong correlation between $\overline{\Delta p}_x$ of them at high $p_T$ for any configuration of e.m. fields generated in heavy ion collisions. The measurement of this correlation should be a strong probe of e.m. fields, where $\overline{\Delta p}_x$ can be extracted utilizing Eq (\ref{v1}) by the measurement of $\Delta v_1(p_T)$. Moreover, as the effect of the Lorentz force on high $p_T$ leptons and charm quarks differs mainly by their different charge, the correlation between charm quarks and leptons from $Z^0$ decay should not only apply to the study of directed flow splitting, but also apply to all other charge dependent observables, such as the spectra ratio, and higher flow splitting of positively and negatively charged particles.

\section{Conclusions}
We have pointed out that the sign and the size of $\Delta v_1$ splitting is not simply proportional to the strength of the e.m field,
see Eq. (\ref{final}) and Eq. (\ref{v1}). Also, even if the determination of electric conductivity cannot be discussed here, our study
shows that  assuming a medium under equilibrium condition $\Delta v_1^D$ becomes positive only for
 $\sigma_{el} \gtrsim 0.05 \,\rm fm^{-1}$ which can be still in agreement with the  lattice QCD estimate. 
We propose for the first time in uRHICs the measurement of the $v_1$ of leptons from $Z^0$ decay to confirm
the electromagnetic origin of the $\Delta v_1^D$ and have a novel constraint on e.m. strength and time evolution. A novel and not expected prediction is that $|\Delta v_1^l|\, < \, |\Delta v_1^D|$ for any $p_T$, see Eq. (\ref{v1}), even if the lepton has a charge $50\%$ larger than the charm quark,  a negligible mass and does not undergo  a damping due to the in-medium strong
interaction, and furthermore they can have an opposite sign at $p_T < 45 \, \rm GeV$, even if both charm quarks and leptons are undergoing the same e.m. field.
This work should trigger the experimental research as well as extended studies that include advancement in understanding
the physics determining the evolution of the electromagnetic fields, a thorough study
as a function of energy and centrality,  and also the extension to the $B^0$ mesons. 
Despite the complexity of the dynamics in ultrarelativistic collisions, we have been able to find out the
key features determining the size of $\Delta v_1$ and we can foretell  the following scaling for $B^0$ mesons $
\frac{\Delta v^B_1(p_T, y_z)}{\Delta v^D_1(p_T, y_z)}\ \simeq 
|\frac{q_b}{q_c}| \frac{d\ln f_b/dp_T}{d\ln f_c/dp_T} $; such a prediction remains
valid if the $\tau_{1} B(\tau_{1})$ remain the same as the charm one, otherwise one has to consider the
impact of the different formation time on $\tau_1$, see Eq.(\ref{final}), and  B meson could provide some new insight
into the very early stage. This certainly requires further detailed studies following up this Letter.

The scope is, however, much wider than the understanding of heavy flavor dynamics and
the generation of the electromagnetic field in the early stage. Providing an independent novel probe of the e.m. field
is a key aspect that can trigger a breakthrough in the new ongoing search for the CME and CVE \cite{Kharzeev:2007jp,Fukushima:2008xe,Kharzeev:2009fn,Jiang:2016wve,Shi:2017cpu,Sun:2018idn}, 
CMW \cite{Kharzeev:2010gd,Burnier:2011bf,Yee:2013cya,Sun:2016nig}, 
and splitting in the $\Lambda$ polarization ~\cite{Becattini:2016gvu,Han:2017hdi,Guo:2019joy}.
Finally, we highlight that if the splitting $\Delta v_1^{D}=v_1(D^0) - v_1 (\overline{D}^0)$ for neutral charged particles
has an electromagnetic origin,
it provides also a direct probe of the existence of the deconfined phase with the charged charm quarks as degrees of freedom, because without the deconfined phase one can expect zero $\Delta v_1^{D}$ due to the absence of the effect of Lorentz force.
This represents an absolutely new and unique probe 
allowing to access the deconfinement as a function of the flavor, a key and open question of the understanding 
of the QCD phase transition \cite{Bazavov:2014yba,Ratti:2018ksb}. 
\\
\section*{ACKNOWLEDGEMENTS}
V.G. and S.P. acknowledge the stimulating discussion with A. Rossi.
The work of  Y.S. is supported by a INFN post-doc fellowship within the national SIM project.
S.P. and V.G. acknowledge the support of the linea di intervento 2-HQCDyn, DFA-Unict.

\end{document}